\documentclass[11pt,preprint]{article}
\usepackage{lineno}
\usepackage[margin=1in]{geometry}
\usepackage{amsmath,amssymb,bm}
\usepackage{graphicx}
\usepackage[numbers,sort&compress]{natbib}
\usepackage[colorlinks=true,allcolors=blue]{hyperref}

\newcommand{\kb}{k_{\rm B}}

\newcommand{\diag}{\operatorname{diag}}
\newcommand{\dd}{\mathrm d}
\newcommand{\grad}{\nabla}
\newcommand{\thetaVec}{\boldsymbol{\theta}}
\newcommand{\x}{\mathbf x}
\newcommand{\z}{\mathbf z}
\newcommand{\uu}{\mathbf u}
\newcommand{\eps}{\boldsymbol\epsilon}
\newcommand{\sig}{\boldsymbol\sigma}
\newcommand{\lag}{\boldsymbol\eta}
\newcommand{\G}{\widehat G}
\newcommand{\Dkl}{D_{\rm KL}}
\newcommand{\Wass}{\mathcal{W}_2}
\newcommand{\Bures}{\mathcal{B}}
\newcommand{\Cc}{\mathcal{C}}

\newcommand{\muthetaval}{0.01}          
\newcommand{\tautrainval}{20}           
\newcommand{\dtval}{0.05}               
\newcommand{\npresval}{3000}            
\newcommand{\Nthetaval}{1510}           

\title{\bfseries Thermodynamic cost of inference and learning\\ in physical neural networks}

\author{
Alexei V. Tkachenko\\[0.3em]
\small Center for Functional Nanomaterials, Brookhaven National Laboratory, Upton, NY 11973, USA\\
\small Correspondence: \texttt{oleksiyt@bnl.gov}
}

\date{}

\begin{document}
\maketitle

\begin{abstract}
How much of the energy consumed by artificial neural networks is set by physics rather than by implementation? For irreversible digital hardware the reference is Landauer's principle, which charges $\kb T\ln2$ per erased bit. We map a generic feedforward network onto a physical Hamiltonian in which each layer relation is an elastic compatibility constraint, and obtain two exact bounds. First, its equilibrium free energy is independent of the input and of every weight and bias, at all temperatures, so quasi-static inference requires no work whatsoever: no thermodynamic cost attaches to computation itself. Second, at finite speed the work exceeds the squared Wasserstein-2 distance between the initial and final thermal states, divided by the protocol duration. Relaxed to an entropic measure of distinguishability, this identifies the cost with the information separating successive inputs, about $\kb T$ per dimension of the widest layer at the fastest usable speed. Learning is fundamentally different: writing the parameters carries an irreducible cost of a few $\kb T$ each that survives the quasi-static limit. The thermodynamic price of a neural network is therefore set by its memory rather than its arithmetic. Simulations confirm both bounds: the inference work saturates the transport bound to within four percent, and accuracy collapses once the dissipated work falls below the thermal scale.
\end{abstract}


The rapid growth of deep neural networks has enabled major advances in natural language processing, computer vision, and molecular biology \citep{vaswani2017attention,he2016deep,jumper2021highly}. These advances have been accompanied by rapidly increasing energy demands for both training and inference \citep{strubell2019energy,LLama,deepseek2025}, motivating a fundamental question: what energetic costs are intrinsic to neural computation, independently of the inefficiencies of current hardware \citep{horowitz2014computing}?

For digital computing the standard benchmark is Landauer's principle \citep{landauer1961,bennett1982thermodynamics,parrondo2015thermodynamics}, which assigns a minimum dissipation $\kb T\ln 2$ to the irreversible erasure of one bit. Landauer's bound applies to logically irreversible operations on metastable information-bearing states; it does not directly constrain continuous analogue computation. In reversible or quasistatic physical systems dissipation can in principle be made arbitrarily small, and the unavoidable cost of finite-speed operation is instead controlled by relaxation, friction, and the geometry of the driven state space \citep{zurek1989thermodynamic,lloyd2000ultimate,sivak2012thermodynamic,still2012thermodynamics,wolpert2019stochastic}.

These considerations are directly relevant to physical neural networks built from optical, electronic, mechanical or quantum components \citep{Shastri2021,Fu2024,Beer2020,markovic2020physics,Wright2022,photonic2023}, which compute through relaxation, equilibration or wave dynamics rather than clocked digital gates. The idea has gained sharp practical form in recent work on thermodynamic computing, where thermal fluctuations are treated as a computational resource rather than a nuisance: harmonic circuits can be made to perform linear algebra \citep{Aifer2024} and have been realized as an eight-cell analogue processor that samples and inverts matrices \citep{melanson2025thermodynamic}, nonlinear thermodynamic circuits act as neurons and can be trained out of equilibrium \citep{whitelam2024nonlinear,whitelam2026training}, a Langevin system can generate structured data from noise with estimated heat dissipation many orders of magnitude below a digital equivalent \citep{whitelam2026generative}, and an all-transistor probabilistic architecture is projected to match GPU performance on an image benchmark using about $10^{4}$ times less energy \citep{jelincic2026efficient}. Such estimates invite a complementary question, which is the one we address: what does thermodynamics actually permit, independently of any particular device? Here we restrict attention to \emph{energy-based} physical neural networks: systems described by a Hamiltonian or free energy, with inference and learning implemented by relaxation under externally imposed constraints. This excludes networks whose operation relies essentially on nonconservative dynamics, but it includes an important family of models at the intersection of statistical physics, neuromorphic computation and physical learning \citep{Hopfield1982,Ackley1985,Liu2021PRX,Arvind2023,dillavou2022demonstration,AndreaPRE24}.

Modern feedforward networks do not naturally fit this framework: their layer-to-layer maps are directed and generally nonreciprocal, whereas conventional energy-based models require reciprocal interactions. Feedforward structure is nevertheless central to deep learning because it enables direct inference and efficient gradient evaluation by backpropagation \citep{rumelhart1986learning}. A thermodynamic description of feedforward networks therefore requires reconciling directed computation with an underlying energy function.

We show that a generic feedforward network can be embedded in a Hamiltonian by treating each layer relation as an elastic compatibility constraint. The resulting energy has a highly degenerate ground-state manifold coinciding with the feedforward map: for any input and any parameter values the corresponding feedforward state has exactly zero energy. This construction separates the thermodynamic roles of inference and training. Inference is an isostatic response along the compatible manifold: in the quasi-static limit it is exactly reversible, and we show that this is not an artefact of the particular embedding but a general property of monostable free-energy machines. Because no real device operates quasi-statically, we then derive the cost of finite-speed operation and evaluate it at the fastest useful rate, when the input is switched on the internal relaxation time $\tau_{\rm inf}$. This is the number that should be compared with hardware, and it is of order $\kb T$ per dimension of the widest layer, so that it scales with the width of the network rather than with its number of parameters. Training, by contrast, clamps both input and target, overconstrains the same free energy, and carries a dissipation that survives the quasi-static limit and scales with both the number of parameters and the size of the training set. We close by comparing all three quantities with the measured energy use of contemporary models and with the Landauer benchmark of a conventional irreversible digital implementation.

\section*{Results}

\subsection*{Hamiltonian embedding and reversible inference}

\begin{figure}[t]
\centering
\includegraphics[width=0.96\linewidth]{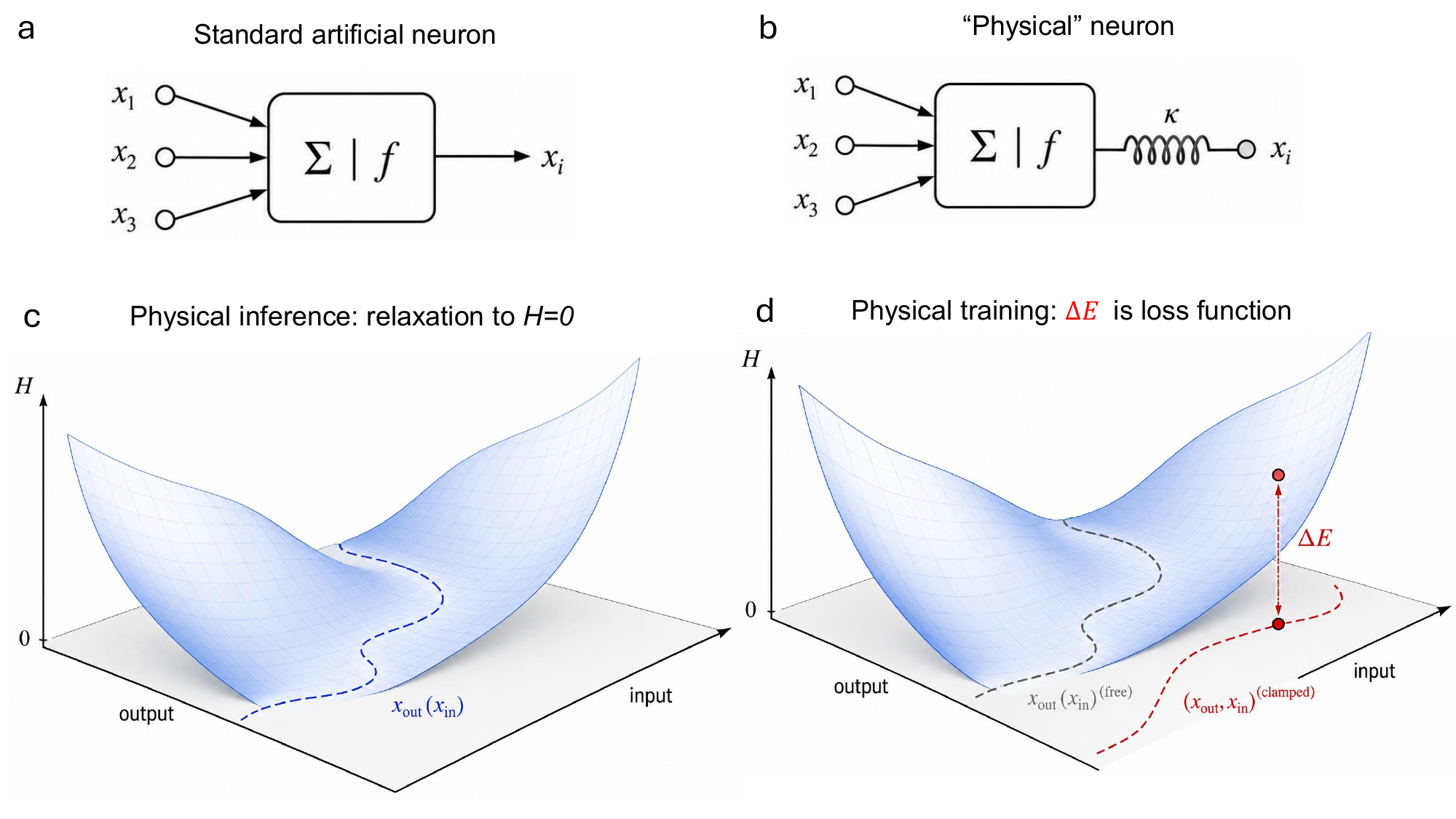}
\caption{\textbf{Elastic embedding of a feedforward neural network.}
\textbf{a,} Standard artificial neuron: inputs are summed and passed through a nonlinear function to produce $x_i$.
\textbf{b,} Physical neuron: the feedforward value is coupled to the output coordinate by an elastic element of stiffness $\kappa$.
\textbf{c,} Inference: the Hamiltonian forms a parabolic canyon with zero-energy floor along the free manifold $x_{\rm out}(x_{\rm in})$.
\textbf{d,} Training: clamping to a target curve displaces the system from the free manifold; the excess energy acts as the physical loss.}
\label{fig:concept}
\end{figure}

Consider a depth-$L$ feedforward network with layer variables $\x_n\in\mathbb R^{d_n}$, input $\x_0=\uu$, output $\x_L=\mathbf y$, and trainable parameters
\begin{equation}
\thetaVec=\{W_n,\mathbf b_n\}_{n=1}^{L}.
\end{equation}
The standard feedforward recursion is
\begin{equation}
\x_n=f_n(\z_n),\qquad \z_n=W_n\x_{n-1}+\mathbf b_n,\qquad n=1,\ldots,L,
\label{eq:feedforward_main}
\end{equation}
where $f_n$ acts componentwise and may include linear, rectified or saturating nonlinearities. Although Eq.~\eqref{eq:feedforward_main} is directed and generally nonreciprocal, it can be embedded in the elastic energy
\begin{equation}
\mathcal H(\x;\thetaVec)=
\frac{\kappa}{2}\sum_{n=1}^{L}
\left\|\x_n-f_n(W_n\x_{n-1}+\mathbf b_n)\right\|^2 .
\label{eq:hamiltonian_main}
\end{equation}
Each layer relation is thus realized by an elastic element of stiffness $\kappa$ that couples the feedforward value to the neuron coordinate (Fig.~\ref{fig:concept}a,b). For any imposed input $\uu$ and any parameters $\thetaVec$, the feedforward state $\x^*(\uu;\thetaVec)$ satisfies $\mathcal H=0$. The Hamiltonian therefore has a highly degenerate ground-state manifold: every input selects a compatible zero-energy state, and the input acts as a boundary constraint fixing the zero modes of the structure.

The stiffness $\kappa$ makes each layer relation a two-sided elastic coupling. The dynamics generated by Eq.~\eqref{eq:hamiltonian_main} is therefore not itself feedforward: forces propagate both downstream and upstream. This is a feature rather than a defect, and is precisely what produces the adjoint recursion of backpropagation as a mechanical force balance in the clamped problem below.

This construction differs from Hopfield networks and Boltzmann machines, where symmetric couplings define an energy landscape and useful states are attractors or equilibrium distributions \citep{Hopfield1982,Ackley1985}. Equation~\eqref{eq:hamiltonian_main} instead assigns a physical free energy to a directed feedforward computation by treating each layer relation as an elastic compatibility condition. A formally identical sum of squared layer-wise mismatches appears as the objective function of predictive coding in computational neuroscience \citep{rao1999predictive,whittington2017approximation,millidge2020predictive}, where its minimization by activity relaxation serves as an inference and learning algorithm approximating backpropagation. Our contribution is therefore not the residual construction itself, but its reading as a physical Hamiltonian: the exact input- and parameter-independence of the resulting partition function, and the finite-time thermodynamic bounds that follow. In the elastic analogy, free inference is an isostatic response: the imposed input determines the internal state without generating strain in the ideal quasistatic limit. The landscape is a parabolic canyon whose floor is the zero-energy manifold, and inference is motion along that floor (Fig.~\ref{fig:concept}c).

We can now ask what inference costs. Start with a statement that holds for any machine of this kind, not just for Eq.~\eqref{eq:hamiltonian_main}. Suppose a device computes by relaxing on a free-energy landscape $F(\x;\uu)$ that is controlled through $\uu$, and suppose that landscape has a single minimum which moves continuously as $\uu$ changes. Driven slowly, the system stays in equilibrium, so the work done is $W=\Delta F$. Since $F$ is a state function, nothing is dissipated: whatever work is spent switching from one input to another is returned on switching back, and the average work over any closed cycle of inputs is zero.

A rate-independent cost appears only if the landscape has several minima that are long-lived on the timescale of operation. Then the system can be left in a state that is not the equilibrium one, and erasing that memory is irreversible. A digital bit is exactly such a pair of long-lived states, which is where Landauer's $\kb T\ln2$ comes from. It is the price of multistable memory, not of computation, and a monostable machine does not pay it. Our network is monostable: for a given input the compatible state is unique, since the energy of Eq.~\eqref{eq:hamiltonian_main} is a positive-definite quadratic form in the deviations from it, as shown below.

For the embedding of Eq.~\eqref{eq:hamiltonian_main} a stronger statement holds. Clamp $\x_0=\uu$ and integrate the remaining coordinates over $\mathbb R^{d_n}$. The change of variables from layer coordinates to residuals $\eps_n=\x_n-f_n(W_n\x_{n-1}+\mathbf b_n)$ is a bijection whose Jacobian is block lower-triangular with unit diagonal blocks, so its determinant is one for any activation functions. The partition function factorizes into $L$ independent Gaussian integrals and
\begin{equation}
F_{\rm free}=\frac{\kb T}{2}\sum_{n=1}^{L}d_n\ln\frac{\kappa}{2\pi \kb T}
\label{eq:free_energy_const}
\end{equation}
does not depend on the input at all, exactly and at every temperature. So $\Delta F=0$ for every pair of inputs, not merely on average around a cycle: the reversible work of a single transition is itself zero, and
\begin{equation}
E_{\rm inf}^{\min}=0 .
\label{eq:Einf_zero}
\end{equation}
Equation~\eqref{eq:Einf_zero} is an idealization in the same sense as Carnot efficiency: it is approached only as the operating speed goes to zero. What follows is the cost of running at finite speed, and in particular at the fastest rate at which the machine still computes.

\subsection*{The cost of finite speed}

At finite speed the network is a stochastic system driven between two thermal states, one for the old input and one for the new. What we show next is that the work of that transition is bounded below by how far apart those two states are, measured as a statistical distance between probability distributions. Two such distances matter here.

The first is the Wasserstein-2 distance $\Wass$ of optimal transport, which measures how far probability mass must be physically moved to turn one distribution into the other. It yields exact minimum-dissipation results valid arbitrarily far from equilibrium \citep{benamou2000computational,jko1998variational,aurell2011optimal,dechant2019thermodynamic,nakazato2021geometrical,vanvu2023thermodynamic,shiraishi2018speed,ito2023geometric}, and has recently been confirmed experimentally \citep{oikawa2025experimental}. The second is the Fisher--Rao distance of information geometry, which measures instead how {\it distinguishable} two distributions are given their own noise. Its infinitesimal form is the Fisher metric $C^{-1}$, and for a mean shift at fixed covariance its finite-displacement form is the Kullback--Leibler divergence $\Dkl$; integrating the Fisher metric along a driving protocol gives the thermodynamic length underlying the near-equilibrium bounds \citep{sivak2012thermodynamic,blaber2023optimal}.

The two answer different questions, and both are needed. Dissipation is a transport quantity, because heat is generated by moving physical coordinates. Computational reliability is a distinguishability quantity, because a result is only usable if it can be told apart from thermal noise. We therefore derive the transport bound first, since it is exact, and then relax it into a Fisher--Rao form expressed through $\Dkl$, which is what sets the accuracy at finite temperature.

Dissipation enters when the input is changed in a finite time $\tau$ and the internal variables lag behind the moving zero-energy manifold. We use overdamped dynamics with scalar mobility $\mu$,
\begin{equation}
\dot\x_n=-\mu\,\grad_{\x_n}\mathcal H+\text{thermal noise},
\qquad n=1,\ldots,L,
\label{eq:langevin_main}
\end{equation}
with noise satisfying the fluctuation--dissipation relation at temperature $T$.

This dynamics has an intrinsic timescale, and it is worth extracting before anything else, because it is what "finite time" must be compared against. Let $\lag_n=\x_n-\x^*_n$ be the lag of layer $n$ behind the instantaneous feedforward state at fixed input, with $\lag_0=0$. Linearizing the residuals gives $\eps_n=\lag_n-J_n\lag_{n-1}$ with $J_n=\diag[f_n'(\z_n)]W_n$ the local Jacobian, or $\eps=A\lag$ with $A$ a block lower-bidiagonal compatibility operator. Then $\mathcal H=\tfrac{\kappa}{2}\lag^TA^TA\lag$ and the fixed-input dynamics is $\dot{\lag}=-\mu\kappa A^TA\lag$. Since $A$ has unit diagonal blocks it is invertible and $A^TA$ is positive definite, which is the monostability invoked above; its smallest eigenvalue sets the slowest relaxation time,
\begin{equation}
\tau_{\rm inf}=\frac{1}{\mu\kappa\,\lambda_{\min}(A^TA)} .
\label{eq:tau_exact_main}
\end{equation}
It is useful to introduce, alongside this exact time, the characteristic diffusive timescale of a depth-$L$ chain,
\begin{equation}
\tau_L\equiv L^2\tau_0 ,
\qquad \tau_0=(\mu\kappa)^{-1} .
\label{eq:tau_inf_main}
\end{equation}
The quadratic dependence on depth is that of diffusion: a locally coupled chain of $L$ layers has no ballistic mode, so a disturbance crosses it the way a random walker crosses a distance, in a time scaling as the square of its length. For a path-like network whose layer Jacobians have singular values close to unity, a near-isometric chain, the smallest singular value of $A$ scales as $L^{-1}$ and the two times are comparable, $\tau_{\rm inf}\sim\tau_L$; departures from unit gain degrade the network on either side, as quantified in the Supplementary Information.

Dissipation in an overdamped system is the Rayleigh functional: the drag force on each coordinate is $\mu^{-1}\dot\x_n$. Minimizing the work over all ways of moving the system between two given thermal states is exactly the optimal-transport problem: by the Benamou--Brenier theorem \citep{benamou2000computational,jko1998variational}, the minimum of the kinetic action over all paths connecting two probability distributions is the squared Wasserstein-2 distance between them, the squared distance that probability mass must be moved \citep{aurell2011optimal,dechant2019thermodynamic,nakazato2021geometrical,vanvu2023thermodynamic}. It gives the exact bound
\begin{equation}
E_{\rm inf}=\frac{1}{\mu}\int_0^{\tau}\sum_n\|\dot\x_n\|^2\dd t\;\ge\;\frac{\Wass^2}{\mu\,\tau}\;=\;\frac{\kappa\tau_0}{\tau}\,\Wass^2 ,
\label{eq:master_bound}
\end{equation}
where $\Wass$ is evaluated between the thermal states associated with the old and new inputs. For thermal endpoint states with means $\x^*_{i},\x^*_{i+1}$ and covariances $\Sigma_i,\Sigma_{i+1}$, the Gaussian form of the Wasserstein distance \citep{givens1984class} splits into a mean shift and a covariance change,
\begin{equation}
\Wass^2=
\sum_{n=1}^{L}\left\|\Delta\x^*_n\right\|^2
\;+\;
\Bures^2(\Sigma_i,\Sigma_{i+1})
\;\ge\;
\sum_{n=1}^{L}\left\|\Delta\x^*_n\right\|^2 ,
\label{eq:gaussian_w2}
\end{equation}
where the first term is a bare Euclidean displacement, and $\Bures$ is the Bures distance between the two covariances. The Bures term is suppressed precisely in the regime where the network computes reliably (Supplementary Information).

The mean-shift form can now be relaxed one step further, into a statement about how distinguishable the computed trajectory is from thermal noise. Within a fixed activation sector the layer maps are linear and the endpoint expressions below are exact; across sector changes the rigorous object is the Fisher length accumulated along the trajectory, of which the endpoint form is the fixed-covariance evaluation (Supplementary Information). For the architecture simulated in this work the interlayer map is linear, and the endpoint form is exact. At each instant define the downstream Green functions $\G_{\ell\to m}=J_m\cdots J_{\ell+1}$ from the local Jacobians introduced above, and let $C_m=(\kb T/\kappa)\,\Cc_m$ with $\Cc_m=\sum_{\ell\le m}\G_{\ell\to m}\G_{\ell\to m}^{T}$ be the instantaneous layer-$m$ covariance. Motion along the manifold satisfies $\Delta\x_m=\G_{\ell\to m}\Delta\x_\ell$, and summing this identity over the $m$ upstream layers and applying Cauchy--Schwarz gives $\sum_{\ell\le m}\|\Delta\x_\ell\|^2\ge m^2\,\Delta\x_m^T\Cc_m^{-1}\Delta\x_m$ (Supplementary Information). We emphasize that nothing here invokes linear response: the transport bound is exact at all driving speeds, and each step of the chain is a statement about the instantaneous linearization at the current operating point, holding at fixed activation pattern, with $\Cc_m$ re-evaluated along the trajectory where activation boundaries are crossed (Supplementary Information). Every layer therefore supplies a valid bound, and we keep the strongest,
\begin{equation}
E_{\rm inf}\;\ge\;\frac{\kappa\tau_0\,\Wass^2}{\tau}\;\ge\;\frac{2\kb T\tau_0}{\tau}\,\max_{1\le m\le L}\;m^2\,\Dkl^{(m)} ,
\label{eq:inference_bound_main}
\end{equation}
where $\Dkl^{(m)}=\tfrac12\langle\Delta\x_m^TC_m^{-1}\Delta\x_m\rangle_{\rm tr}$ is the mean-shift Kullback--Leibler divergence between the initial and final thermal distributions of layer $m$ \citep{Kullback1951} and $\langle\cdots\rangle_{\rm tr}$ averages over input transitions. Taking the maximum rather than evaluating at a chosen layer matters: the layer that binds is the one carrying the most information relative to its own noise, and it need not be the output, whose width may be unrepresentative of the network.

Let $d$ be the width of the network. Define the inference fidelity by
\begin{equation}
\Gamma_{\rm inf}\equiv\frac{2}{L^2 d}\max_{1\le m\le L}m^2\,\Dkl^{(m)} .
\label{eq:gamma_inf_def}
\end{equation}
This quantity has the meaning of an effective squared signal-to-noise ratio, and a conservative one. Writing $m^*$ for the layer attaining the maximum, $\Gamma_{\rm inf}=(m^*/L)^2(d_{m^*}/d)\times2\Dkl^{(m^*)}/d_{m^*}$: since $L$ bounds $m^*$ from above and $d$ bounds the layer width $d_{m^*}$, both prefactors are at most one, so $\Gamma_{\rm inf}$ is a lower estimate of the per-coordinate signal-to-noise ratio $2\Dkl^{(m^*)}/d_{m^*}$ actually realized at the most informative layer. It is a fidelity in the operational sense: it measures how far the computed state moves relative to the noise it must be told apart from, and $\Gamma_{\rm inf}\gtrsim1$ guarantees that the true signal-to-noise ratio also exceeds one. Substituting Eq.~\eqref{eq:gamma_inf_def} into Eq.~\eqref{eq:inference_bound_main} is then an identity-level rewriting, the factor $L^2$ combining with $\tau_0$ to form exactly the diffusive timescale of Eq.~\eqref{eq:tau_inf_main},
\begin{equation}
E_{\rm inf}\;\ge\;
d\,\kb T\,\Gamma_{\rm inf}\,\frac{\tau_L}{\tau} .
\label{eq:inference_scaling_main}
\end{equation}
For a gain-normalized network $\tau_L\simeq\tau_{\rm inf}$, so the price of speed is measured against the machine's own relaxation. Reliable finite-temperature inference requires $\Gamma_{\rm inf}$ large enough to separate the computed state from equilibrium fluctuations, and faster operation raises the price linearly in $\tau_L/\tau$. Notably, no material parameters survive: $\mu$ and $\kappa$ enter only through $\tau_0$, so the bound depends on a fidelity, a speed ratio and a coordinate count alone, which is what allows it to be applied across photonic, electronic and mechanical platforms.

Each inequality in the chain leading to Eq.~\eqref{eq:inference_scaling_main} is close to saturation in the regime where the machine is meant to operate. The transport bound is attained by constant-speed driving; input-only control reaches it in the slow-driving limit for linear response, the case of the network simulated below (Supplementary Information). Dropping the Bures term costs nothing here, since the endpoint covariances coincide exactly for this architecture, and in general its relative weight is $\sim1/(L^2\Gamma_{\rm inf})$. The Cauchy--Schwarz step is tight when the per-layer displacements are comparable, the gain-normalized regime, in which $m^*\simeq L$ and $d_{m^*}\simeq d$ so that $\Gamma_{\rm inf}$ approaches the realized signal-to-noise ratio itself. The bound is therefore not a loose chain of relaxations but an estimate accurate up to order-unity factors for a well-conditioned network, degrading only where the network itself does; the simulations below confirm this directly, with the slow-driving work saturating the transport bound to within four percent (Supplementary Information).

Equation~\eqref{eq:inference_scaling_main} interpolates between the reversible limit, Eq.~\eqref{eq:Einf_zero}, and the fastest rate at which the machine still computes. Driving faster than $\tau_{\rm inf}$ does not reduce the cost of a computation; it fails to complete one, as the simulations below confirm. The operationally meaningful benchmark is therefore $\tau\simeq\tau_{\rm inf}$, which for a gain-normalized network equals $\tau_L$ and gives
\begin{equation}
E_{\rm inf}\big|_{\tau=\tau_{\rm inf}}\simeq d\,\Gamma_{\rm inf}\,\kb T .
\label{eq:benchmark}
\end{equation}
Equation~\eqref{eq:benchmark} simplifies further at the fastest speed the machine can actually be run. Reliable inference requires a fidelity $\Gamma_{\rm inf}\gtrsim10$, and the same layer gains that raise $\Gamma_{\rm inf}$ also slow the relaxation, so that the exact relaxation time and the diffusive scale are related by $\tau_{\rm inf}\sim\Gamma_{\rm inf}\tau_L$. Setting the operating time to $\tau=\tau_{\rm inf}$ in Eq.~\eqref{eq:inference_scaling_main} makes the fidelity cancel,
\begin{equation}
E_{\rm inf}\big|_{\tau=\tau_{\rm inf}}\;\gtrsim\;d\,\kb T ,
\label{eq:benchmark_simple}
\end{equation}
about $\kb T$ per dimension of the widest layer the computation must resolve. The estimate is useful provided $\tau_{\rm inf}$ remains practically short; it is the width of the network, not its parameter count, that sets the price.

\subsection*{Physical training and its irreducible cost}

Training is implemented by clamping both the input and the target output. For a data pair $\alpha=(\uu_\alpha,\mathbf t_\alpha)$ we impose $\x_0=\uu_\alpha$, $\x_L=\mathbf t_\alpha$ and let the hidden variables equilibrate, defining the constrained free energy $F_\alpha(\thetaVec)$ as the usual logarithm of the restricted partition function over $\x_1,\ldots,\x_{L-1}$. Clamping displaces the system off the zero-energy manifold, and the excess energy is the loss (Fig.~\ref{fig:concept}d). At low temperature $F_\alpha$ reduces to the minimum elastic strain energy compatible with the imposed data.

Equation~\eqref{eq:free_energy_const} is what makes this work as a learning rule. Besides being independent of the input, $F_{\rm free}$ contains no weights or biases, so the free network exerts no force on the parameters whatsoever. The clamped free energy is therefore the loss by itself, and no subtraction of free and clamped energies is required, in contrast to equilibrium-propagation-type schemes.

Training is thus the overconstrained counterpart of isostatic inference: input and target clamps are generally incompatible with the current feedforward map, so they generate strain, and the parameter gradients are the corresponding internal stresses. Define
\begin{equation}
\eps_n=\x_n-f_n(\z_n),
\qquad
\sig_n=\kappa\,[f_n(\z_n)-\x_n]\circ f_n'(\z_n),
\label{eq:stress_def_main}
\end{equation}
with $\circ$ componentwise multiplication. By the envelope theorem, at the constrained minimum,
\begin{equation}
\frac{\partial F_\alpha}{\partial \mathbf b_n}=\sig_n,
\qquad
\frac{\partial F_\alpha}{\partial W_n}=\sig_n\x_{n-1}^{T},
\label{eq:parameter_gradients_main}
\end{equation}
with thermal averages inserted at finite temperature. Force balance on a hidden layer gives
\begin{equation}
\sig_n=(W_{n+1}^{T}\sig_{n+1})\circ f_n'(\z_n),
\qquad n=L-1,\ldots,1,
\label{eq:backprop_force_main}
\end{equation}
which is the adjoint recursion of backpropagation. Backpropagation therefore appears as mechanical equilibrium in the overconstrained physical network.

This mechanism is related in spirit to equilibrium propagation, physics-aware training, Hamiltonian echo backpropagation and other proposals for learning in physical systems \citep{scellier2017equilibrium,Wright2022,Science2023,lopezpastor2023self,Physical_PNAS2024,momeni2024arxive,Liu2021PRX,Arvind2023,dillavou2022demonstration}. The special feature of Eq.~\eqref{eq:hamiltonian_main} is that the feedforward manifold has exactly zero energy for any parameters, so training can be driven by simple parameter annealing in the ordinary constrained free energy rather than by a separately engineered learning rule.

The intrinsic training cost has two contributions of very different character. The first is the finite-rate work required to change the clamped input--target pair, to which the inference analysis applies verbatim with the clamped trajectory in place of the free one. If $N_{\rm pres}$ data presentations are switched over a time $\tau_{\rm feed}$,
\begin{equation}
E_{\rm feed}\ge
\frac{1}{\mu\,\tau_{\rm feed}}\sum_{p=1}^{N_{\rm pres}}\left[\sum_n\|\Delta\x^*_n\|^2\right]_p
\;\gtrsim\;
\kb T\,N_{\rm pres}\,d\,\Gamma_{\rm inf}
\frac{\tau_L}{\tau_{\rm feed}} .
\label{eq:feed_scaling_main}
\end{equation}
Like the inference cost, this term vanishes as the presentation slows; it is expected to dominate in high-throughput devices.

The second contribution does not vanish. Let
\begin{equation}
\overline F(\thetaVec)=\langle F_\alpha(\thetaVec)\rangle_\alpha,
\qquad
\mathbf g_\alpha=\grad_\theta F_\alpha,
\qquad
\overline{\mathbf g}=\langle\mathbf g_\alpha\rangle_\alpha,
\end{equation}
where $\langle\cdots\rangle_\alpha$ averages over the training distribution. For mobility matrix $M_\theta$, the parameter-space dissipation rate is $\langle \mathbf g_\alpha^TM_\theta\mathbf g_\alpha\rangle_\alpha$, whereas the useful decrease of the averaged loss is $\overline{\mathbf g}^{T}M_\theta\overline{\mathbf g}$. The stochastic-force incoherence
\begin{equation}
\mathcal R_\theta=
\frac{
\left\langle \mathbf g_\alpha^TM_\theta\mathbf g_\alpha\right\rangle_\alpha-
\overline{\mathbf g}^{T}M_\theta\overline{\mathbf g}
}{
\overline{\mathbf g}^{T}M_\theta\overline{\mathbf g}
}
\label{eq:rtheta_main}
\end{equation}
measures the fraction of parameter motion that does not contribute to average descent, and the residual heat released while the averaged free energy drops by $\dd F_{\rm drop}=-\dd\overline F>0$ is
\begin{equation}
\dd Q_\theta=(1+\mathcal R_\theta)\,\dd F_{\rm drop} .
\label{eq:residual_diss_main}
\end{equation}

Thermal fluctuations in parameter space set the scale on which learning is possible at all. Near a minimum of the averaged loss, equipartition distributes $\tfrac12\kb T$ over each soft parameter direction, so the landscape is thermally blurred over an energy range $N_\theta\kb T$ and the learned state is resolvable only if the achieved decrease of the averaged free energy exceeds it. With the training fidelity
\begin{equation}
\Gamma_{\rm train}=\frac{\Delta\overline F}{N_\theta\kb T},
\label{eq:gamma_train_main}
\end{equation}
robust learning therefore requires $\Gamma_{\rm train}\gtrsim1$ as an order-of-magnitude resolvability criterion (with $N_\theta$ standing for the number of effectively stiff parameter directions, for which the nominal count is the natural scale), and combining this with Eq.~\eqref{eq:residual_diss_main} gives the parameter-annealing cost $E_\theta\gtrsim\Gamma_{\rm train}(1+\overline{\mathcal R}_\theta)N_\theta\kb T$. Adding the data-presentation term,
\begin{equation}
E_{\rm train}\gtrsim
\kb T\left[
N_{\rm pres}\,d\,\Gamma_{\rm inf}\frac{\tau_L}{\tau_{\rm feed}}
+
\Gamma_{\rm train}(1+\overline{\mathcal R}_\theta)N_\theta
\right].
\label{eq:training_scaling_main}
\end{equation}
The two terms are qualitatively different. The first is the finite-rate cost of moving the clamped coordinates between examples; it carries a protocol time and can be made arbitrarily small by slow presentation. The second carries no protocol time: it survives the quasi-static limit and scales with the number of stored parameters, so within the passive dynamics considered it cannot be removed by slowing down. Dataset size enters the second term only through $\overline{\mathcal R}_\theta$, which reflects the heterogeneity of the per-example parameter forces. Equation~\eqref{eq:training_scaling_main} is far weaker than the information-theoretic bounds obtained from the mutual information between true and inferred labels \citep{Goldt2017,GoldtPRL}, but unlike them it references the physical complexity of the machine that does the learning.

\subsection*{Numerical tests}

\begin{figure}[h!]
\centering
\includegraphics[width=1\linewidth]{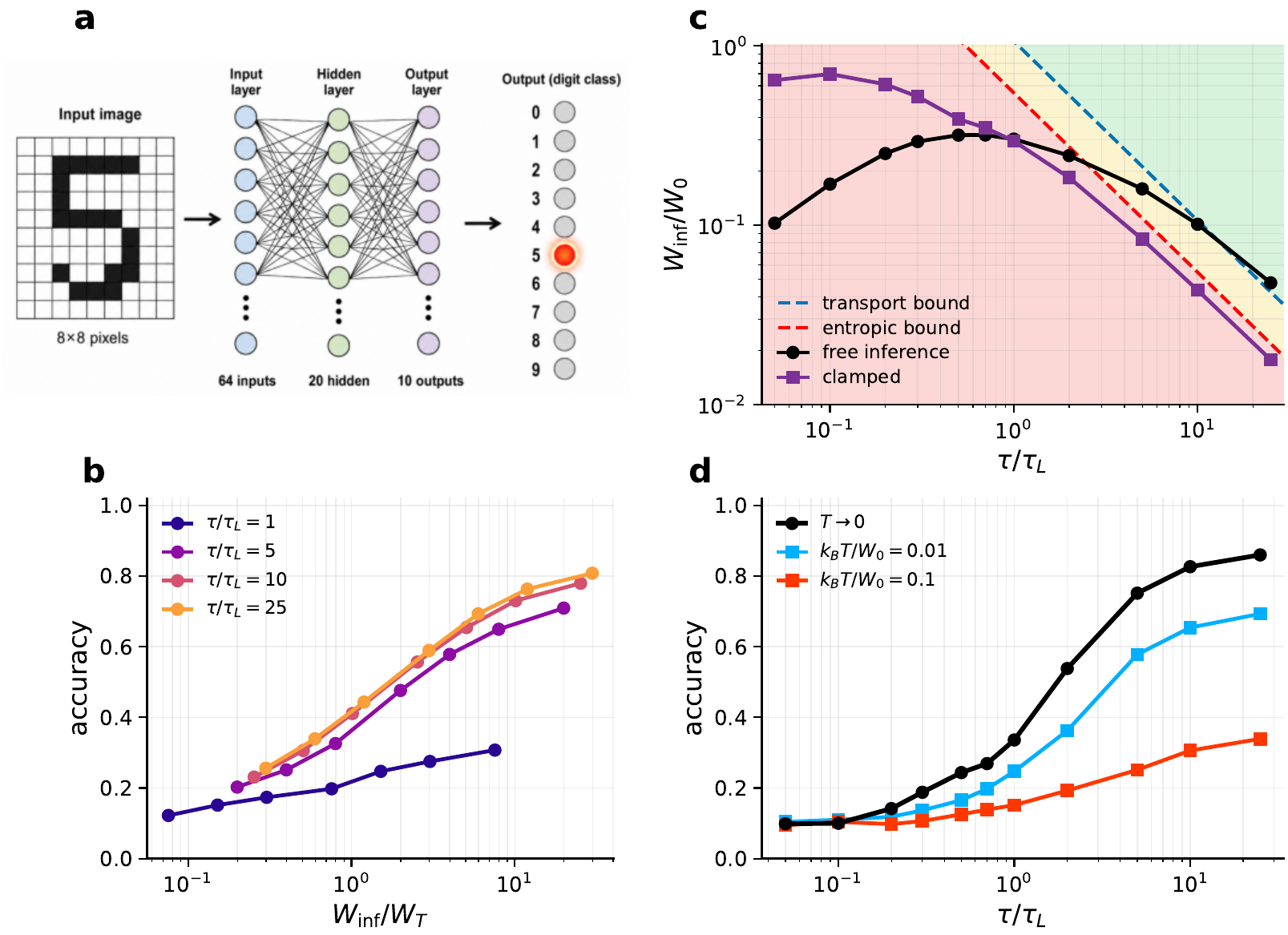}
\caption{
\textbf{Inference work under input interpolation in a physical neural network.}
\textbf{a,} Schematic of the physical network, in which feedforward inference is embedded as relaxation to a zero-energy manifold.
\textbf{b,} Test accuracy versus thermally normalized inference work $W_{\rm inf}/W_T$, with $W_T=\kb T\,d\,\tau_L/\tau$, showing a crossover from chance-level accuracy to deterministic accuracy near $W_{\rm inf}/W_T\simeq1$.
\textbf{c,} Input-control work versus switching time, compared with the transport bound (blue) and its information-geometric relaxation (orange), both evaluated per transition between the feedforward endpoints and falling as $\tau^{-1}$ on these axes. The shading marks the regions below the entropic bound (red), between the two bounds (yellow) and above the transport bound (green). Free inference crosses from one region to the next as the transition completes, saturating the transport bound to within four percent on the slow branch. Because the bounds refer to completed transitions between ideal feedforward endpoints, which the persistent physical state reaches only as $\tau$ grows, the measured work at $\tau\lesssim\tau_L$ lies below them: it reflects transitions that were never completed rather than cheaper computation, and coincides with the accuracy collapse in \textbf{d}. The clamped protocol moves between clamped equilibria, whose transport distance is smaller, and is therefore not bounded by the lines shown.
\textbf{d,} Accuracy versus $\tau/\tau_L$ in the deterministic limit $T\to0$ and at finite temperature, showing degradation when the dissipated inference work becomes comparable to the thermal distinguishability scale.
}
\label{fig:inference}
\end{figure}

We tested the framework in a minimal physical neural network trained on the handwritten-digits data set distributed with \texttt{scikit-learn}, with architecture $64\to20\to10$ (Fig.~\ref{fig:inference}a). Inputs were externally imposed while the hidden and output variables were physical degrees of freedom. The hidden transfer function was rectified linear and the output layer linear, so that
\begin{equation}
\mathcal H=\frac{1}{2}\left\|\mathbf h-\operatorname{ReLU}(W_1\uu+\mathbf b_1)\right\|^2
+\frac{1}{2}\left\|\mathbf y-W_2\mathbf h-\mathbf b_2\right\|^2 ,
\label{eq:numerics_energy_main}
\end{equation}
the two-layer realization of Eq.~\eqref{eq:hamiltonian_main}. The simulations were designed not to optimize classification accuracy but to test how work, relaxation and thermal reliability follow from the geometry of the zero-energy manifold.

\paragraph{Inference.} The input was switched linearly between consecutive test examples over a time $\tau$ while the hidden and output variables relaxed, with the physical state persistent across transitions. The measured input-control work approached $E_{\rm inf}\propto\tau^{-1}$ in the slow-driving regime, as predicted (Fig.~\ref{fig:inference}c). The work is non-monotonic and peaks near $\tau\sim\tau_L$: for $\tau\ll\tau_L$ the persistent internal state cannot track the input at all, so little work is exchanged through the input control and no computation is performed, and the classification accuracy is correspondingly at chance (Fig.~\ref{fig:inference}d). Low work at high speed therefore reflects a transition that was never completed, not a cheaper computation, and the bounds of Eqs.~\eqref{eq:master_bound} and \eqref{eq:inference_bound_main}, which refer to completed transitions, should be read against the slow-driving branch. At finite temperature, accuracy was controlled by the ratio of dissipated work to the thermal distinguishability scale
\begin{equation}
E_T(\tau)=\kb T\,d\,\frac{\tau_L}{\tau},
\end{equation}
with $d=20$ the network width of Eq.~\eqref{eq:gamma_inf_def}. The collapse variable $w=E_{\rm inf}/E_T(\tau)$ is a thermally normalized work, bounded below by $\Gamma_{\rm inf}$ through Eq.~\eqref{eq:inference_scaling_main}. Plotted this way, data taken at four protocol times and seven temperatures collapse onto a common crossover from chance-level classification to the deterministic physical-inference accuracy near $w\simeq1$ (Fig.~\ref{fig:inference}b), consistent with the fidelity threshold $\Gamma_{\rm inf}\simeq1$. The one protocol time that does not collapse, $\tau=\tau_L$, is the one for which the transition is only marginally completed.

\begin{figure}[h!]
\centering
\includegraphics[width=0.63\linewidth]{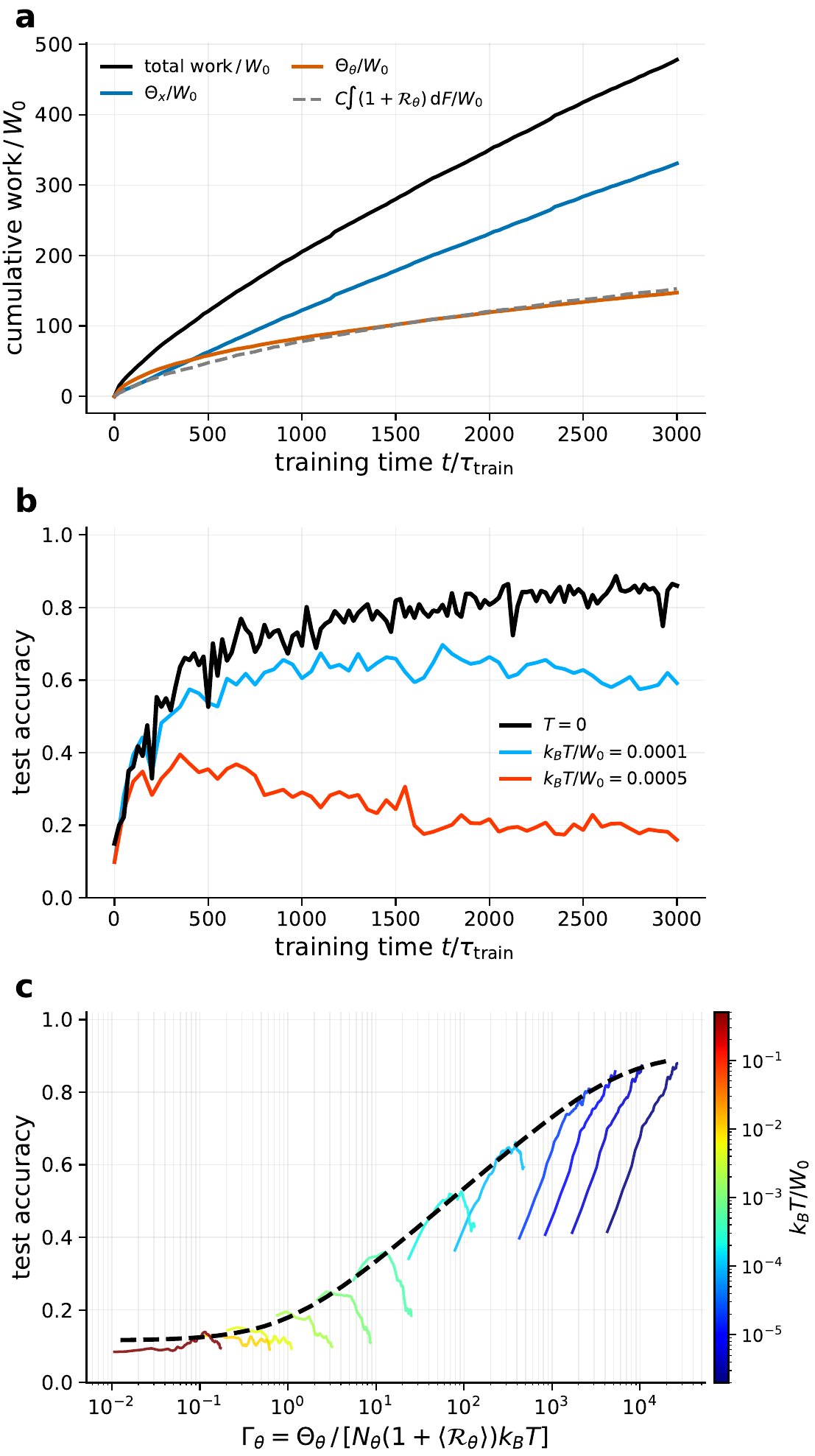}
\caption{
\textbf{Intrinsic thermodynamic cost of physical training.}
\textbf{a,} Cumulative training work, split into fast-variable dissipation $\Theta_x$ and parameter-annealing dissipation $\Theta_\theta$; the latter follows the fitted $C\int(1+\mathcal R_\theta)\dd F_{\rm drop}$.
\textbf{b,} Test accuracy during training at $T=0$ and at finite parameter temperature, which limits accuracy and can degrade it after initial learning.
\textbf{c,} Accuracy versus the training fidelity $\Gamma_\theta$, Eq.~\eqref{eq:gamma_theta_main}, for twelve temperatures (one colour each; hotter trajectories end at smaller $\Gamma_\theta$). Dashed: logarithmic fit to the upper-left envelope, Eq.~\eqref{eq:meth_envelope} ($R^2=0.995$).
}
\label{fig:training}
\end{figure}

\paragraph{Training.} Training used the same Hamiltonian with the output clamped to the target during each presentation, input and target switched linearly between consecutive training examples in randomized order, and the hidden state persistent throughout. Because the output is clamped, the only fast physical coordinate is the hidden state, and the measured cumulative work separates into a fast-variable contribution $\Theta_x=\int\|\dot{\mathbf h}\|^2\dd t$ and a residual parameter-annealing contribution $\Theta_\theta$. For the representative mobility $\mu_\theta=\muthetaval/\tau_0$, $\Theta_x$ exceeded $\Theta_\theta$ by roughly a factor of two over $\npresval$ presentations (Fig.~\ref{fig:training}a). The onset of learning near $\Gamma_\theta\simeq1$ is a direct test of the rate-independent term of Eq.~\eqref{eq:training_scaling_main}.

The smaller residual term tested the intrinsic cost of learning. It followed the predicted form $\Theta_\theta\simeq C\int(1+\mathcal R_\theta)\,\dd F_{\rm drop}$ with a single order-unity prefactor obtained by least squares through the origin, $C=2.96$ with $R^2=0.973$. This supports the interpretation of $\mathcal R_\theta$ as stochastic-force incoherence: the coherent component of the data-dependent force lowers the average constrained free energy, while the incoherent component produces heat.

Finite-temperature parameter dynamics were characterized by the measured fidelity
\begin{equation}
\Gamma_\theta=\frac{\Theta_\theta}{N_\theta(1+\langle\mathcal R_\theta\rangle)\kb T},
\label{eq:gamma_theta_main}
\end{equation}
which is the dissipative counterpart of $\Gamma_{\rm train}$, the two being related by Eq.~\eqref{eq:residual_diss_main} up to the prefactor $C$. Learning fails entirely for $\Gamma_\theta\lesssim1$, as required by Eq.~\eqref{eq:gamma_train_main}. Reaching high accuracy, however, requires $\Gamma_\theta$ well above this threshold: the empirical envelope of Fig.~\ref{fig:training}c follows the logarithmic fit of Eq.~\eqref{eq:meth_envelope}, whose large-$\Gamma_\theta$ limit gives the accuracy deficit $\Delta A\simeq|A|\Gamma_0/\Gamma_\theta\approx5.6\times10^{2}/\Gamma_\theta$.

\section*{Discussion}

The construction developed here assigns a physical free energy to a directed feedforward network by treating each layer relation as an elastic compatibility constraint. The computational advantage of feedforward architectures is preserved, and a statistical-mechanical description is restored. Three results follow, and they answer different questions.

The first concerns what is possible in principle. The equilibrium free energy does not depend on the input or on any parameter, exactly and at every temperature, so slow inference connects states of equal free energy and dissipates nothing. The general reason is that the landscape has a single minimum for each input: Landauer's bound applies to memory built from long-lived alternative states, not to computation as such, and a machine that never creates such states never pays it. This is why an analogue network can in principle undercut a digital one by an amount that has nothing to do with better engineering.

The second concerns what is possible in practice, since no device runs infinitely slowly. Moving the network in a finite time $\tau$ costs at least the optimal-transport distance between its endpoints divided by $\tau$, and relaxing that statement gives the signal-to-noise form $E_{\rm inf}\ge\kb Td\,\Gamma_{\rm inf}\tau_L/\tau$, which also governs accuracy at finite temperature. Running faster than the relaxation time does not buy cheaper computation, it fails to complete one; at the meaningful operating point $\tau\simeq\tau_{\rm inf}$ the fidelity cancels against the relaxation time it also controls, leaving about $\kb T$ per dimension of the widest layer, Eq.~\eqref{eq:benchmark_simple}. That the count is dimensions and not operations is the essential point: a digital implementation pays an erasure cost per weight and so scales with the parameter count $N_\theta$, while the physical machine performs those multiplications through the geometry of its own zero-energy manifold and pays only for the coordinates it moves, scaling with the width $d$.

The third concerns learning, and it goes the other way. Clamping input and target together drives the system from an isostatic to an overconstrained state: the resulting strain is the loss, the parameter forces are internal gradient signals, and the backpropagation recursion is mechanical force balance, so learning is not an algorithm imposed on the physics but a consequence of relaxation in the same landscape that performs inference. Its cost splits into a data-presentation term that vanishes with slow presentation and a parameter-annealing term $\Gamma_{\rm train}(1+\overline{\mathcal R}_\theta)N_\theta\kb T$ that does not. The asymmetry with inference is structural: any sufficiently slow protocol makes inference free, whereas under the passive annealing that implements learning physically, no operating speed does. The contrast with digital training is correspondingly sharper than for inference. A digital run pays its erasure cost afresh at every presentation, so its benchmark carries the full dataset size $D$, while the physical cost of a few $\kb T$ per parameter is incurred once, for the parameter values that are finally stored.

\medskip

The digital costs follow from counting erasures rather than moved coordinates: with $\Gamma_{\rm dig}$ bits discarded per operation, Landauer's principle gives $2N_\theta\Gamma_{\rm dig}\kb T\ln2$ per token for inference and $6N_\theta D\,\Gamma_{\rm dig}\kb T\ln2$ for a training run of $D$ presentations (Methods). Table~\ref{tab:compare} evaluates these against the physical bounds for a $65$-billion-parameter dense model and a $671$-billion-parameter mixture-of-experts model trained on $D=1.5\times10^{13}$ tokens \citep{LLama,deepseek2025}, at $T=300$~K with $\Gamma_{\rm dig}=16$ and $d\sim10^{6}$; only orders of magnitude are meaningful.

\begin{table}[t]
\centering
\begin{tabular}{l c c c c}
\hline\hline
 & Current & Landauer benchmark & \multicolumn{2}{c}{This work}\\
\cline{4-5}
Energy per token (J) & hardware & (irreversible 16-bit) & slow limit & $\tau  \sim \tau_{\rm inf}$\\
\hline
Inference (65B dense) & $10^{0}$ & $10^{-8}$ & $0$ & $10^{-14}$\\
Training (671B sparse) & $10^{-1}$ & $10^{-8}$ & $10^{-22}$ & $10^{-14}$\\
\hline\hline
\end{tabular}
\caption{\textbf{Thermodynamic bounds against practice.} Energy per token for contemporary large models, compared with the Landauer benchmark for a conventional irreversible 16-bit digital implementation and with the bounds derived here, at $T=300$~K. Inference entries are Eqs.~\eqref{eq:Einf_zero} and \eqref{eq:benchmark}. Training entries are the two terms of Eq.~\eqref{eq:training_scaling_main}: in the slow limit only the parameter-annealing term survives, amounting to $N_\theta\kb T$ spread over the whole training set, while at $\tau=\tau_{\rm inf}$ the cost of moving the clamped coordinates dominates. All entries are order-of-magnitude estimates; measured values are taken from Refs.~\citep{LLama,deepseek2025} and published accelerator specifications, and the evaluation of every entry is detailed in the Supplementary Information.}
\label{tab:compare}
\end{table}

It is also worth calibrating against hardware that already exploits physics rather than logic. The all-transistor probabilistic architecture of Ref.~\citep{jelincic2026efficient} dissipates about $2$~fJ per conditional update of a single sampling cell. Taken at face value, this sits roughly $5\times10^{4}$ above the per-coordinate bound $\Gamma_{\rm inf}\kb T$ of Eq.~\eqref{eq:benchmark}, against the $10^{7}$ by which digital accelerators exceed their own Landauer benchmark: physics-based machines have already closed most of the gap on a logarithmic scale, and the remaining margin is what the present bounds quantify.

Two features stand out. Almost all of the inference margin is a matter of counting rather than of $\kb T$ being small, since the physical cost tracks the $10^{6}$ coordinates that move while the digital benchmark tracks the $10^{11}$ operations performed. Training keeps an irreducible term even when run arbitrarily slowly, but that term is minute in absolute terms because it is a one-time cost of order $N_\theta\kb T$ amortized over a very large dataset. At full speed both phases are limited by the same thing, the work of moving physical coordinates between successive inputs.

These figures are intrinsic costs for an ideal substrate rather than device budgets. Real machines also pay for data conversion, control, measurement, memory, imperfect materials, error correction and the physical realization of the dense nonlinear couplings themselves, and those costs dominate today. The mapping also assumes the affine-plus-activation structure of Eq.~\eqref{eq:feedforward_main}, and does not address attention or normalization. In particular, because $F_{\rm free}$ is flat in the parameter directions, retaining the learned weights after training requires a confining or latching potential $U(\thetaVec)$ whose write, hold and reset costs lie outside the present accounting: the bounds derived here price the acquisition of the parameters, not their storage. Two further points favour digital implementations and should be kept in view: sparse architectures activate only a small fraction of their parameters per token and pay the digital cost accordingly, whereas physical annealing acts on every trainable parameter; and a trained digital model can be copied at negligible cost, which has no analogue in a physically annealed network.

\medskip

In conclusion, we have represented a generic feedforward network as a physical free-energy landscape whose equilibrium value, Eq.~\eqref{eq:free_energy_const}, is independent of the input and of all weights and biases. As a result, quasistatic inference requires no net work: the system moves between equilibrium states of equal free energy, and any cost that remains in the slow-operation limit is associated with changing the parameters rather than evaluating the network. At finite speed, the required work is bounded below by the optimal-transport result, Eq.~\eqref{eq:master_bound}, and by its information-theoretic relaxation, Eq.~\eqref{eq:inference_scaling_main}. Our simulations saturate the transport bound to within four percent of the theoretical minimum, with the relaxed form lying about a factor of two below it. They also show that accuracy deteriorates once the work becomes comparable to thermal fluctuations. At the fastest useful operating speed, the fidelity cancels against the relaxation time it also controls, and the cost is about $\kb T$ per dimension of the widest layer, Eq.~\eqref{eq:benchmark_simple}: it scales with the width of the network rather than with its parameter count, and lies about five orders of magnitude below the Landauer benchmark for an equivalent irreversible digital calculation and thirteen orders below present hardware estimates (Table~\ref{tab:compare}).

Learning has a different thermodynamic character because it requires the parameters themselves to change. The corresponding cost, given by Eq.~\eqref{eq:training_scaling_main}, does not vanish when the protocol is slowed, amounting to a few $\kb T$ per parameter for the whole of training; a digital run, by contrast, pays its erasure cost at every presentation and so carries the full dataset size as an additional factor. Numerically, this residual dissipation is proportional to the useful decrease in free energy, with a correction set by the variability of the parameter forces across training examples. For the task studied, the empirical envelope further indicates that the energy required to reduce the remaining accuracy error increases inversely with that error. Thus, inference is limited primarily by the finite-rate motion of the physical state, whereas learning is limited by the energetic cost of establishing the parameter configuration. In this sense, the thermodynamic cost of a neural network is associated more with memory than with arithmetic.

\section*{Methods}

\subsection*{Numerical model}

The numerical tests used the handwritten-digits data set distributed with \texttt{scikit-learn}. Pixel values were divided by 16 so that all input components lay in $[0,1]$, and the data were split into training and test sets by a stratified split with test fraction 0.35 and fixed seed, giving 1168 training and 629 test examples. The architecture was $64\to20\to10$ with a rectified-linear hidden layer and a linear output layer; targets were one-hot vectors. Weights were initialized as $W_1\sim\mathcal N(0,2/d_0)$ and $W_2\sim\mathcal N(0,1/d_1)$ with $d_0=64$, $d_1=20$, and biases at zero. The number of trainable parameters was $N_\theta=20\times64+20+10\times20+10=\Nthetaval$.

The physical degrees of freedom were the hidden state $\mathbf h$ and, during free inference, the output state $\mathbf y$. For input $\uu$ define $\mathbf a(\uu)=\operatorname{ReLU}(W_1\uu+\mathbf b_1)$. The free-inference energy was
\begin{equation}
\mathcal H=\tfrac12\left\|\mathbf h-\mathbf a(\uu)\right\|^2+\tfrac12\left\|\mathbf y-W_2\mathbf h-\mathbf b_2\right\|^2 ,
\label{eq:meth_H_free}
\end{equation}
and for training the output was clamped to the target $\mathbf t$,
\begin{equation}
\mathcal H_{\rm cl}=\tfrac12\left\|\mathbf h-\mathbf a(\uu)\right\|^2+\tfrac12\left\|\mathbf t-W_2\mathbf h-\mathbf b_2\right\|^2 .
\label{eq:meth_H_cl}
\end{equation}
Mobility and stiffness were set to unity in simulation units, so $\tau_0=(\mu\kappa)^{-1}=1$ and all times are quoted in units of $\tau_0$. All time axes are normalized by the diffusive timescale of Eq.~\eqref{eq:tau_inf_main}, $\tau_L=L^2\tau_0=4\tau_0$ for the $L=2$ physical layers simulated. For the trained model, $\sigma_{\max}(W_2)=3.21$ gives $\lambda_{\min}(A^TA)=0.0816$ and hence the exact relaxation time $\tau_{\rm inf}=12.2\,\tau_0=3.1\,\tau_L$, an order-unity multiple of the diffusive scale as expected. The spectrum of $A^TA$ is available in closed form for this architecture, one reciprocal pair of modes per singular value of $W_2$; it is derived and displayed in the Supplementary Information, where the slow modes are seen to form a band of which $\tau_{\rm inf}$ is the conservative representative. All times reported in the figures and text are in units of $\tau_L$; the exact $\tau_{\rm inf}$ enters only through this ratio, and none of the qualitative features depend on the distinction.

\subsection*{Inference simulations}

Inference was simulated by overdamped relaxation of $\mathbf h$ and $\mathbf y$,
\begin{equation}
\dot{\mathbf h}=-(\mathbf h-\mathbf a)+W_2^T(\mathbf y-W_2\mathbf h-\mathbf b_2),
\qquad
\dot{\mathbf y}=-(\mathbf y-W_2\mathbf h-\mathbf b_2),
\label{eq:meth_infer_dyn}
\end{equation}
using the parameters of the trained $T=0$ model, while the input was switched linearly, $\uu(t)=(1-s)\uu_i+s\uu_{i+1}$ with $s=t/\tau$, between consecutive test examples in their stored order. The physical state was persistent across transitions and was not reset. Protocol times $\tau/\tau_L\in[0.05,25]$ were scanned, with the number of integration steps set to $\max(20,\lceil\tau/0.1\rceil)$. The classification decision was read from $\mathbf y$ at the end of each switching window, with no additional relaxation.

The input-control work was accumulated by the trapezoidal rule,
\begin{equation}
E_{\rm inf}=\int\frac{\partial\mathcal H}{\partial\uu}\cdot \dd\uu,
\qquad
\frac{\partial\mathcal H}{\partial\uu}=-W_1^T\left[(\mathbf h-\mathbf a)\circ\operatorname{ReLU}'(W_1\uu+\mathbf b_1)\right],
\label{eq:meth_work}
\end{equation}
reported per transition and normalized by the hidden-state jump scale evaluated on the same sequence,
\begin{equation}
E_0=\tfrac12\left\langle\left\|\mathbf a(\uu_i)-\mathbf a(\uu_{i-1})\right\|^2\right\rangle_i .
\label{eq:meth_E0}
\end{equation}
Because $\mathcal H$ vanishes at both ends of a completed transition, this control work equals the dissipated heat; for $\tau\ll\tau_L$ the transition is not completed and the measured control work is not the full dissipation. A clamped protocol, in which the correct one-hot target was additionally imposed and switched over the same window, was run for comparison, with the output-clamp work included.

Finite-temperature inference added fluctuation--dissipation noise of variance $2\kb T\,\dd t$ per step to both $\mathbf h$ and $\mathbf y$, with $\kb T$ quoted relative to $E_0$. Accuracies were averaged over five independent noise realizations, and the collapse of Fig.~\ref{fig:inference}b was constructed by plotting the mean accuracy against $E_{\rm inf}/E_T(\tau)$ with $E_T(\tau)=\kb T d\,\tau_L/\tau$ and $d=20$, using the deterministic $E_{\rm inf}$ at the matching $\tau$.

Each physical layer carries its own residual fluctuation, so $\Cc_m\succeq I$ and the layer covariances are positive definite whatever the activation pattern; no active-subspace restriction is needed, and $d_m$ is the full layer width. For this architecture the rectified nonlinearity acts on the clamped input and therefore propagates no thermal fluctuations at all: the layer covariances are $C_1=(\kb T/\kappa)I$ and $C_2=(\kb T/\kappa)(I+W_2W_2^T)$, both independent of the input. Since they are the same at both ends of every transition, the covariance contribution to the transport bound vanishes identically for this model and Eq.~\eqref{eq:master_bound} holds with no relaxation at all. The maximization in Eq.~\eqref{eq:inference_bound_main} is performed over both physical layers, and the network width in Eq.~\eqref{eq:gamma_inf_def} is taken as the hidden width, $d=20$.

\subsection*{Training simulations}

Training used $\mathcal H_{\rm cl}$ with both input and target switched linearly between consecutive training examples, presented in an order reshuffled each epoch. The hidden state was persistent between presentations and obeyed
\begin{equation}
\dot{\mathbf h}=-(\mathbf h-\mathbf a)+W_2^T\mathbf r_2,
\qquad
\mathbf r_2=\mathbf t-W_2\mathbf h-\mathbf b_2 ,
\label{eq:meth_train_dyn}
\end{equation}
and with $\mathbf r_1=(\mathbf h-\mathbf a)\circ\operatorname{ReLU}'(W_1\uu+\mathbf b_1)$ the parameter dynamics was
\begin{equation}
\dot W_1=\mu_\theta \mathbf r_1\uu^T,\quad
\dot{\mathbf b}_1=\mu_\theta\mathbf r_1,\quad
\dot W_2=\mu_\theta \mathbf r_2\mathbf h^T,\quad
\dot{\mathbf b}_2=\mu_\theta\mathbf r_2 .
\label{eq:meth_param_dyn}
\end{equation}
The deterministic simulations used $\tau_{\rm train}=\tautrainval\,\tau_0$ per presentation, $\Delta t=\dtval\,\tau_0$, $\mu_\theta=\muthetaval/\tau_0$ and $\npresval$ presentations (about 2.6 epochs), with accuracy checkpoints every 25 presentations.

The fast-variable dissipation was $\Theta_x=\int\|\dot{\mathbf h}\|^2\dd t$ and the residual parameter-annealing dissipation $\Theta_\theta=\int\mu_\theta\|\mathbf g_\theta\|^2\dd t$, with $\mathbf g_\theta$ the instantaneous physical parameter force assembled from $\mathbf r_1$ and $\mathbf r_2$. Both were accumulated at every integration step. The averaged constrained free energy $\overline F$ was estimated by the window-averaged instantaneous clamped energy $\mathcal H_{\rm cl}$ along the trajectory, and the stochastic-force incoherence was evaluated in consecutive non-overlapping windows of duration $25\tau_0$ as
\begin{equation}
\mathcal R_\theta=
\frac{\left\langle\|\mathbf g_\theta\|^2\right\rangle_{\rm win}-\left\|\left\langle\mathbf g_\theta\right\rangle_{\rm win}\right\|^2}
{\left\|\left\langle\mathbf g_\theta\right\rangle_{\rm win}\right\|^2},
\label{eq:meth_Rtheta}
\end{equation}
which is Eq.~\eqref{eq:rtheta_main} for isotropic mobility $M_\theta=\mu_\theta I$. The prediction $\Theta_\theta\simeq C\int(1+\mathcal R_\theta)\dd F_{\rm drop}$ was tested by a single-parameter least-squares fit through the origin, discarding the first two per cent of the cumulative integral.

Finite-temperature training added a Langevin kick in parameter space, accumulated over each presentation and applied as an independent Gaussian increment of standard deviation $\sigma_{\rm pres}=(2\mu_\theta\kb T\tau_{\rm eff})^{1/2}$ to every parameter. For $\kb T/E_0$ above $10^{-4}$, $\tau_{\rm eff}$ and the integration step were reduced by a common stability factor $\min\{[(\kb T/E_0)/10^{-4}]^{1/2},30\}$, leaving the number of steps per presentation unchanged; the effective presentation time is therefore shorter at the highest temperatures. Twelve values of $\kb T/E_0$ between $2\times10^{-6}$ and $0.5$ were run, each for two independent realizations differing in both initialization and noise, with checkpoints every 50 presentations. Trajectories were averaged over realizations and smoothed with a centred seven-point moving average in presentation number, edge points without full support being discarded.

\subsection*{Envelope construction and fit}

The smoothed trajectories were binned into 36 logarithmic bins in $\Gamma_\theta$; the largest smoothed accuracy in each bin was taken and a running maximum applied, giving the upper-left envelope. The envelope was then fitted by
\begin{equation}
A_{\rm test}(\Gamma_\theta)=a_{\max}+A\ln\!\left(1+\frac{\Gamma_0}{\Gamma_\theta+1}\right),
\label{eq:meth_envelope}
\end{equation}
a three-parameter fit in which $\Gamma_0$ is scanned on a logarithmic grid and $A$, $a_{\max}$ are obtained by linear least squares at each $\Gamma_0$; none of the three is fixed in advance. The best fit gives $A=-0.0911$, $\Gamma_0=6.15\times10^{3}$, $a_{\max}=0.9105$ with $R^2=0.995$. The coefficient $A$ is negative because accuracy increases with $\Gamma_\theta$ while the logarithm decreases; the text quotes $|A|$, and the figure displays the fitted curve without parameter annotation. The low-$\Gamma_\theta$ plateau $a_{\max}+A\ln(1+\Gamma_0)\approx0.12$ is an outcome of the fit rather than an imposed chance level, and the large-$\Gamma_\theta$ limit of Eq.~\eqref{eq:meth_envelope} is the inverse law $\Delta A\simeq|A|\Gamma_0/\Gamma_\theta$ quoted in the main text.

\subsection*{Digital benchmark}

The digital costs quoted in Table~\ref{tab:compare} count erasures rather than moved coordinates. Autoregressive inference performs $2N_\theta$ floating-point operations per token and training $6N_\theta$ per token, with $N_\theta$ the number of parameters active per token \citep{kaplan2020scaling}. Assigning $\Gamma_{\rm dig}$ erased bits to each operation, so that $\Gamma_{\rm dig}$ measures an effective arithmetic fidelity in the same sense that $\Gamma_{\rm inf}$ measures a physical one, Landauer's principle gives
\begin{equation}
E^{\rm dig}_{\rm inf}=2N_\theta\,\Gamma_{\rm dig}\,\kb T\ln2
\label{eq:meth_dig_inf}
\end{equation}
per token for inference and
\begin{equation}
E^{\rm dig}_{\rm train}=6N_\theta D\,\Gamma_{\rm dig}\,\kb T\ln2
\label{eq:meth_dig_train}
\end{equation}
for a complete training run over $D$ presentations. Table~\ref{tab:compare} uses $\Gamma_{\rm dig}=16$, appropriate to 16-bit arithmetic, and quotes the training figure per token.

Two differences of scaling separate these expressions from the physical bounds of Eqs.~\eqref{eq:benchmark_simple} and \eqref{eq:training_scaling_main}, and they are more consequential than any prefactor. For inference, Eq.~\eqref{eq:meth_dig_inf} scales with the parameter count $N_\theta$ while the physical cost scales with the layer width $d$; the two differ by the ratio of synapses to neurons, which is $10^{4}$--$10^{5}$ in contemporary architectures, because the physical machine performs its multiplications through the geometry of the zero-energy manifold rather than as operations that must each be paid for. For training, Eq.~\eqref{eq:meth_dig_train} carries in addition the full number of presentations $D$, since a digital run erases afresh for every example, whereas the quasi-static physical cost $\Gamma_{\rm train}(1+\overline{\mathcal R}_\theta)N_\theta\kb T$ is incurred once, for the parameter values that are finally stored. With $D\sim10^{13}$ this second factor is by far the larger of the two.

\subsection*{Transport and information-geometric bounds}

The measured work was compared quantitatively with the transport and entropic bounds evaluated per transition from the trained-model geometry, and the finite-temperature training protocol was rechecked at fixed physical presentation time with five independent seeds per temperature; both analyses are reported in the Supplementary Information.

\section*{Data availability}

The numerical examples use the public handwritten-digits data set available through \texttt{scikit-learn}.

\section*{Code availability}

The simulation and analysis code (Jupyter notebook), together with the source data for the figures, is publicly available at \url{https://github.com/atkachen00/Thermodynamic_Bound}.

\section*{Acknowledgements}

This research was performed and used the computational resources at the Center for Functional Nanomaterials, a U.S. DOE Office of Science User Facility, Brookhaven National Laboratory, under Contract No. DE-SC0012704.

\section*{Author contributions}

AT conceived the theoretical framework, performed the analysis and simulations, and wrote the manuscript. AI tools (Claude and ChatGPT) have been used for verification and refinement of the theory, code generation, and manuscript editing. 

\section*{Competing interests}

The author declares no competing interests.

\bibliographystyle{unsrtnat}
\bibliography{main}

\end{document}